\documentclass[11pt,a4paper]{article}
\pdfoutput=1
\usepackage{jheppub}
\usepackage{latexsym}
\usepackage{multirow}
\usepackage{color}
\usepackage[usenames,dvipsnames,table]{xcolor}
\usepackage{graphicx}
\usepackage{epsfig} 
\usepackage{epsf}   
\usepackage{dcolumn}
\usepackage{bm}
\usepackage{dcolumn}
\usepackage{textcomp}
\usepackage{float}
\usepackage{subfig}
\usepackage{hypcap}
\usepackage[]{hyperref}
\usepackage{makecell}
\usepackage{color}
\usepackage{pifont}
\usepackage{appendix}
\usepackage{amsmath}
\usepackage{multirow,bigdelim}  
\usepackage{lineno}
\usepackage[normalem]{ulem}
\graphicspath{{fig/}}

\newcommand{\comment}[1]{}
\usepackage{amssymb}
\hypersetup{
  bookmarks=true,         
  unicode=false,          
  pdftoolbar=true,        
 pdfmenubar=true,        
 pdffitwindow=true,   
 pdfstartview={FitH},    
 pdfsubject={scalar NSI},   
 pdfnewwindow=true,      
 pdfcreator={RevTeX},
 colorlinks=true,     
 linkcolor=red,         
 citecolor=blue,        
 filecolor=black,     
 urlcolor=blue,           
  }

\preprint{}

	\title{Exploring the effects of Scalar Non Standard Interactions on the CP violation sensitivity at DUNE}
	
	\author[a]{Abinash Medhi,}
	\author[b]{Debajyoti Dutta}
	\author[a]{and Moon Moon Devi}
	
	\affiliation[a]{Department of Physics, Tezpur University, Napaam, Sonitpur, Assam-784028, India}
	\affiliation[b]{Department of Physics, Assam Don Bosco University, Kamarkuchi, Sonapur, Assam-782402, India}

	\emailAdd{amedhi@tezu.ernet.in}
	\emailAdd{debajyoti.dutta@dbuniversity.ac.in}
	\emailAdd{devimm@tezu.ernet.in}
	
	\date{\today}
	
	\abstract{The Neutrino oscillations have provided an excellent opportunity to study new-physics beyond the Standard Model, popularly known as BSM. The unknown couplings involving neutrinos, termed non-standard interactions (NSI), may appear as `new-physics' in different neutrino experiments. The neutrino NSI offers significant effects on neutrino oscillations and CP-sensitivity, which may be probed in various neutrino experiments. The idea of neutrinos coupling with a scalar has evolved recently and looks promising. The effects of scalar NSI may appear as a perturbation to the neutrino mass matrix in the neutrino Hamiltonian. It modifies the neutrino mass matrix and may provide a direct possibility of probing neutrino mass models. As the scalar NSI affects the neutrino mass matrix in the Hamiltonian, its effect is energy independent. Moreover, the matter effects due to scalar NSI scales linearly with the matter density.
		
		In this work, we have performed a model-independent study of the effects of scalar NSI at long baseline neutrino experiments, taking DUNE as a case study. We have performed such a thorough study for DUNE for the first time. Various neutrino parameters may get affected due to the inclusion of scalar NSI as it modifies the effective mass matrix of neutrinos. We have explored the impact of scalar NSI in neutrino oscillations and its impact on the measurements of various mixing parameters. We have probed the effects of scalar NSI on different oscillation channels relevant to the experiment. We have also explored the impact of various possible elements in the scalar NSI term on the CP-violation sensitivity at DUNE.
	}
	
	\arxivnumber{2111.12943}
	
\begin{document}
	\maketitle

\section{Introduction}\label{sec:introduction}

The well-established formalism of neutrino oscillations \cite{Super-Kamiokande:1998kpq} have provided one of the first clear evidences of physics beyond the Standard Model (BSM). To explain the phenomena of neutrino oscillations, which confirms neutrinos are not massless, needs an extension of the Standard Model (SM) despite its unprecedented success. As we go beyond SM to explain the neutrino mixing and associated phenomena, many BSM models suggest additional interactions which are generally termed as the non-standard interactions (NSI). The NSIs may impact the production, propagation, and detection of neutrinos in different neutrino experiments and thus necessitate a thorough understanding of the possible impacts. The study of scalar NSI has been a growing field to explore the possibility of new interactions in various neutrino experiments. In this paper, we explore the impact of a general scalar NSI on the neutrino mixing and for the first time, we have thoroughly studied its effects on the CP-violation sensitivity at the proposed long-baseline neutrino experiment DUNE \cite{Abi:2020loh}. 

In the standard interaction scenario, neutrinos may interact with matter via charged-current (CC) and/or neutral-current (NC) interactions mediating a $W^{\pm}$ and/or $Z$ bosons respectively. The idea of neutrino interaction with the matter was initially proposed in \cite{Wolfenstein:1977ue}, where a new matter potential term in the neutrino Hamiltonian due to neutrino matter interactions was introduced. It was proposed that neutrino interactions with matter appear as a matter potential term in neutrino Hamiltonian. Later a few works \cite{MSW1,MSW2,Mikheyev:1985zog} showed that the neutrino mixing can resonate to a maximal value for some particular values of neutrino energy times the matter density. This phenomenon has been termed as the Mikheyev-Smirnov-Wolfenstein (MSW) effect \cite{MSW3}, which successfully explains the results of various experiments \cite{Bahcall:1964gx,Davis:1964hf,deSalas:2017kay,SNO:2002tuh}. The idea of the neutrino matter effect was later explored widely and it has been well established by different experiments. In the same paper \cite{Wolfenstein:1977ue}, where the idea of neutrino matter interactions was proposed, the idea of neutrino NSIs was also indicated using vector boson as mediators of the new non-standard interactions.

 The study of NSI is a well-motivated phenomenological approach to probe new physics beyond the Standard Model. The idea of NSI was introduced with a generally parameterized vector and axial-vector currents \cite{Wolfenstein:1977ue}, and later explored in a large number of studies \cite{Miranda:2015dra, Farzan:2017xzy, Biggio:2009nt, Babu:2019mfe, Ohlsson:2012kf}. The general formalism for vector NSI that has been widely used in the literature is shown in the following, 
 
 \begin{equation}
 V_{NSI} = V_{\mathrm{CC}}
     \begin{bmatrix}
 \varepsilon_{ee}  & \varepsilon_{e\mu}      & \varepsilon_{e\tau}   \\
\varepsilon_{e\mu}^*  & \varepsilon_{\mu\mu}    & \varepsilon_{\mu\tau} \\
\varepsilon_{e\tau}^* & \varepsilon_{\mu\tau}^* & \varepsilon_{\tau\tau}
\end{bmatrix},
 \end{equation}
 
 \noindent where, $V_{CC}$ = $\pm \sqrt 2 G_F n_e$ comes due to neutrino matter effects and $\varepsilon_{\alpha\beta}$ represents the strengths of the vector NSI.
 
 In the reference \cite{Capozzi:2019iqn}, the authors have discussed how one flavour changing NSI element ($\epsilon_{e\tau}$) can significantly affect the mass ordering sensitivities of NO$\nu$A \cite{NOvA:2004blv,NOvA:2016kwd} and T2K \cite{T2K:2014xyt}. This can completely wash out the current $\sim$ 2.4$\sigma$ indication of NO$\nu$A and T2K in favour of the normal hierarchy (NH) of neutrino mass ordering. In \cite{Agarwalla:2016fkh}, the authors showed that the discovery potential of the octant of $\theta_{23}$ of DUNE gets considerably affected in presence of off-diagonal ($\epsilon_{e\mu}$ and $\epsilon_{e\tau}$) vector NSI elements. In presence of these NSI elements the transition probability $P_{\mu e}$ acquires an extra new interference term, which creates a degeneracy in the measurement of $\delta_{CP}$. In the reference \cite{Deepthi:2017gxg}, the authors have explored how diagonal and off-diagonal elements can severely affect the sensitivity of determination of $\delta_{CP}$ at DUNE. Apart from DUNE, the CP sensitivity of other long-baseline (LBL) neutrino experiments like NO$\nu$A, T2K, and proposed T2HK \cite{Hyper-KamiokandeWorkingGroup:2014czz} may also be severely get affected in presence of NSI \cite{PhysRevD.94.013014}. The mass ordering sensitivities of different LBL experiments in presence of NSI have been widely explored in this study~\cite{Masud:2016nuj}. For different studies on the impact of NSI elements on the sensitivities of LBL experiments may be referred to \cite{Friedland:2012tq,Coelho:2012bp,Rahman:2015vqa,Coloma:2015kiu,deGouvea:2015ndi,Liao:2016hsa,Forero:2016cmb,Huitu:2016bmb,Bakhti:2016prn,Kumar:2021lrn,Agarwalla:2015cta,Agarwalla:2014bsa,Agarwalla:2012wf,Blennow:2016etl,Blennow:2015nxa,Deepthi:2016erc,Masud:2021ves,Soumya:2019kto,Masud:2018pig,Masud:2017kdi,Masud:2016bvp,Masud:2015xva,Ge:2016dlx,Fukasawa:2016lew,Chatterjee:2021wac, Dey:2018yht}. A global status on the sensitivity of cp-violation and neutrino mass ordering in presence of NSI in LBL neutrino experiments has been summarised in \cite{Esteban:2019lfo,Coloma:2019mbs}. The study of NSI opens up the potentiality of using neutrino oscillation experiments to explore new-physics scenarios beyond the SM. Such a new interaction leads to a rich phenomenology in both scattering experiments and neutrino oscillation experiments. Various experimental data are being analyzed to see the bounds on such effects \cite{Choubey:2015xha,Khatun:2019tad,Chatterjee:2014gxa,Super-Kamiokande:2011dam, Davidson:2003ha,Choubey:2014iia,Denton:2018xmq,Farzan:2015hkd,Farzan:2015doa,Esmaili:2013fva,Khan:2021wzy, Du:2021idh,Liu:2020emq,Chatterjee:2020kkm,Denton:2020uda,Babu:2020nna,Flores:2020lji,Farzan:2019xor,Pandey:2019apj, Dey:2020fbx}. 
 
 The formulation of possible non-standard couplings of neutrinos to a scalar field has been explored recently \cite{Ge:2018uhz, Yang:2018yvk, Khan:2019jvr}. This type of scalar interaction appears as a correction to the neutrino mass term \cite{Babu:2019iml} and it can have different phenomenological consequences than that of the vector-mediated NSI. Various studies are done to investigate the effects of these scalar NSI elements taking different astrophysical and cosmological constraints also with terrestrial and space-based experimental constraints \cite{Babu:2019iml,Venzor:2020ova}. The presence of scalar NSI has been used to explain the existing Borexino data \cite{Ge:2018uhz}.  Also, as the scalar NSI appears as an addition to the neutrino mass matrix, its impact on neutrino mass models is highly interesting and promising. These scalar couplings may also have an impact on the measurement of different neutrino oscillation parameters in various neutrino oscillation experiments. As the effects of scalar NSI is directly proportional to the environmental matter density and hence it makes long-baseline neutrino experiments one of the suitable candidates to probe its effects. 
 
 In this work, we have performed a model-independent study of the effects of scalar NSI parameters on the long-baseline neutrino experiments, taking DUNE as a case study. It is one of the first thorough studies of such scalar NSI effects at DUNE. We have found that scalar NSI may have a significant impact on the measurements of the $\delta_{CP}$ phase at DUNE. We start with a general formulation for scalar NSI as a matrix and probe its effects element-wise. The details of the formalism of scalar NSI are extensively discussed in the subsequent sections. We have used the diagonal NSI parameters for this work. We observe that the effects of scalar NSI are mostly significant around the oscillation maxima. We also notice the occurrence of various degeneracies in determining $\delta_{CP}$ in presence of the scalar NSI elements. Following that, we have further checked how NSI can impact the cp-violation measurement at DUNE. We show that the effects of diagonal scalar NSI parameters are notably significant at DUNE. We show that for some chosen values of scalar NSI parameters the experiment's sensitivities get enhanced. In addition, we have studied the CP precision measurement potential of DUNE in presence of scalar NSI. We show that the capability of the experiment to constrain $\delta_{CP}$ is significantly affected by the inclusion of these NSI parameters. For certain values of the scalar NSI parameters the precision measurement capability of the experiment reduces, while, for some other NSI parameter values, the capability gets enhanced. Hence, constraining these NSI parameters is extremely crucial for $\delta_{CP}$ sensitivities at DUNE.

We organize the paper as follows: we discuss the formalism of scalar NSI in section \ref{sec:scalar NSI}. In section \ref{sec:Methodology}, we discuss the effect of scalar NSI on the oscillation probabilities in presence of scalar NSI and the details of the simulation methodology used in the analysis. In section \ref{sec:results}, we present the results of the study and finally we summarise our findings in section \ref{sec:conclusion}. 

\section{Scalar NSI}\label{sec:scalar NSI}

 The experiments \cite{Super-Kamiokande:2004orf,KamLAND:2004mhv,MINOS:2008kxu,MINOS:2011neo} covering a wide range of baselines and energies established the theory of neutrino oscillations, i.e.,  neutrinos can change their flavours due to their non-zero masses \cite{Maki:1962mu}. The data from such experiments prove that the neutrino flavours ($\nu_e, \nu_\mu, \nu_\tau$) which are the superposition of their mass states $\nu_1,\nu_2,\nu_3$ with masses $m_1,m_2,m_3$ respectively. The mixing of these flavour states and mass eigenstates are governed by a 3 $\times$ 3 matrix called Pontecorvo-Maki-Nakagawa-Sakata (PMNS) matrix, $\mathcal{U}$ \cite{Pontecorvo:1957cp,Pontecorvo:1957qd,Pontecorvo:1967fh,Maki:1962mu},

\begin{eqnarray}
{\mathcal U}^{} &=& \left(
\begin{array}{ccc}
1   & 0 & 0 \\  0 & c_{23}  & s_{23}   \\ 
0 & -s_{23} & c_{23} \\
\end{array} 
\right)   
\left(
\begin{array}{ccc}
c_{13}  &  0 &  s_{13} e^{- i \delta_{CP}}\\ 0 & 1   &  0 \\ 
-s_{13} e^{i \delta_{CP}} & 0 & c_{13} \\
\end{array} 
\right)  \left(
\begin{array}{ccc}
c_{12}  & s_{12} & 0 \\ 
-s_{12} & c_{12} &  0 \\ 0 &  0 & 1  \\ 
\end{array} 
\right)  \ ,
\label{u}
\end{eqnarray}\\

\noindent where, $s_{ij}=\sin {\theta_{ij}}, c_{ij}=\cos \theta_{ij}$ and $\delta_{CP}$ is the Dirac-type CP phase. This is called the PDG parameterization \cite{ParticleDataGroup:2020ssz} of the PMNS matrix. In addition, if neutrinos are considered as Majorana particles then there may also come two additional phases called Majorana phases. However, these phases do not affect the neutrino oscillations as they can come as a common phase in the neutrino Hamiltonian. We are going to use this parameterization of the PMNS matrix throughout the study.

In the Standard Model interactions scenario, the neutrinos interact with matter via weak interactions only, via a $W^{\pm}$ or Z boson mediator~\cite{Linder:2005fc}. The expression of effective Lagrangian for these interactions is given by~\cite{Wolfenstein:1977ue, Nieves:2003in, Nishi:2004st}, \\

\begin{equation}
\mathcal L^{\rm eff}_{\rm cc}
=
- \frac {4 G_F}{\sqrt 2}
\left[
\overline{\nu_e}(p_3) \gamma_\mu P_L \nu_e(p_2)
\right]
\left[
\overline e(p_1) \gamma^\mu P_L e(p_4)
\right],
\label{eq:Leff}
\end{equation}\\

\noindent where, $ P_L$ and $P_R$ are left and right chiral projection operators respectively, with $P_L = (1 - \gamma_5)/2$ and $P_R=(1+ \gamma_5)/2$),
$p_{i}$'s are momentum of incoming and outgoing states and 
 $G_F$ is the Fermi constant.

 Usually, the neutrino matter effects come from the forward scattering of neutrinos, considering zero momentum transfer between initial and final states. These effects appear as matter potentials in the neutrino Hamiltonian viz. $V_{\rm CC} = \pm \sqrt 2 G_F n_e$ and $V_{\rm NC} = - \frac{G_F n_n}{\sqrt 2} $. Here, $V_{\rm CC}$ and $V_{\rm NC}$ are the matter potentials due to CC and NC matter interactions of neutrinos with the matter. Here, the positive sign arises due to interactions of neutrinos with matter while the negative sign arises due to interactions of antineutrinos. Note that the matter potential due to NC interactions ($V_{\rm NC}$) does not affect the neutrino oscillations since it just appears as a common phase in the neutrino Hamiltonian. The effective Hamiltonian ($\mathcal H_{matter}$) for neutrino oscillations in the matter thus can be written as \cite{Bilenky:1987ty},\\
 
\begin{equation}
\mathcal H_{matter}
\approx
E_\nu
+ \frac { M M ^\dagger}{2 E_\nu}
\pm V_{\rm SI} \,,
\label{eq:matter_H}
\end{equation}\\

\noindent where, $E_\nu$ is the neutrino energy, M is the mass matrix of neutrinos, and $V_{\rm SI}$ is the matter potential due to neutrino matter effects. We emphasize here again that the positive/negative sign of the `$V_{SI}$' term arises due to neutrino/antineutrino modes. The neutrino mass matrix $M$ in flavour basis is given by $\mathcal{U}D_\nu\mathcal{U}^\dagger$, where $D_\nu$ is the diagonal mass matrix of neutrinos i.e. $D_\nu$ $\equiv$ diag($m_1, m_2, m_3$). The simplified effective Hamiltonian ($\mathcal{H_{\rm eff}}$) for neutrino oscillations in matter may be obtained as,

\begin{equation}
\mathcal{H_{\rm eff}} = E_\nu + \frac{1}{2E_\nu} \, \mathcal{U} {\rm diag}(0, \Delta m^2_{21}, \Delta m^2_{31}) \mathcal{U}^\dag + {\rm diag} (V_{\rm CC}, 0 , 0)\,,
\label{eq:matter_H2}
\end{equation}
where,
$\Delta m^2_{ij} \equiv m_i^2 - m_j^2$ are the neutrino mass-squared differences.
The quantity $V_{\rm CC} \equiv \pm \sqrt{2} G_F n_e$ is the effective matter
potential due to the coherent elastic forward scattering of neutrinos
with electrons in the matter through the SM gauge boson $W$.

Although the vector-mediated NSI has been well explored, a non-standard effect may also arise due to other factors. Since neutrinos can couple with a scalar (Higgs boson) with non-zero vacuum expectation values to generate its mass, the coupling of neutrinos with a scalar is an interesting possibility. The effective Lagrangian for such a typical scalar NSI can be framed as~\cite{Ge:2018uhz,Babu:2019iml}, \\

\begin{align}
{\cal L}_{\rm eff}^{\rm S} \ = \ \frac{y_f y_{\alpha\beta}}{m_\phi^2}(\bar{\nu}_\alpha(p_3) \nu_\beta(p_2))(\bar{f}(p_1)f(p_4)) \,, 
\label{eq:nsi_L}
\end{align}\\
where, \\
\noindent $\alpha$, $\beta$ refer to the neutrino flavors e, $\mu$, $\tau$,\\ 
\noindent $f$ = e, u, d indicate the matter fermions, (e: electron, u:
up-quark, d: down-quark),\\
\noindent $\bar{f}$ is for corresponding anti fermions, \\
 \noindent $y_{\alpha\beta}$ is the Yukawa couplings of the neutrinos with the scalar mediator $\phi$, \\
  \noindent$y_f$ is the Yukawa coupling of $\phi$ with $f$, and, \\ \noindent $m_\phi$ is the mass of the scalar mediator $\phi$. \\

From eq.~\ref{eq:nsi_L} we see that the effective Lagrangian is composed of Yukawa terms and as a result of this, it can not be converted into vector currents. As a result, the scalar NSI will not appear as a contribution to the matter potential term in the neutrino Hamiltonian. Instead, it may appear as a medium-dependent perturbation to the neutrino mass term~\cite{Ge:2018uhz}. In this paper, we shall use the suffix ``$SNSI$'' to represent the quantities which include the effects of the scalar NSI.

The corresponding Dirac equation incorporating the new scalar interactions can be simplified \cite{Ge:2018uhz} to,\\ 

\begin{equation}
  \bar \nu_\beta
\left[
  i \partial_\mu \gamma^\mu
+
\left(
  M_{\beta \alpha}
+ \frac {\sum_f n_f y_f y_{\alpha \beta}}{m^2_\phi}
\right)
\right] \nu_\alpha
=
  0 \,,
\end{equation}\\

\noindent with $n_f$ being the number density of the environmental fermions. We see that the contribution from scalar NSI in the Dirac equation is present with the mass term. The effective Hamiltonian \cite{Parke:2019vbs,Akhmedov:2004ny,Babu:2019iml}, taking into account the scalar NSI effects,  will be a modified form of eq.~\ref{eq:matter_H2} as given below, \\

\begin{equation}
\mathcal H_{\rm SNSI}
\approx
E_\nu
+ \frac { M_{\rm eff}  M_{\rm eff}^\dagger}{2 E_\nu}
\pm V_{\rm SI} \,.
\label{eq:Hs}
\end{equation}\\

\noindent Where, $M_{\rm eff}$ is the effective mass matrix that includes both the regular mass matrix $M$ and the contribution from the scalar NSI, $M_{SNSI}  \equiv \sum_f n_f y_f y_{\alpha\beta} / m^2_\phi$ and may be written as,

\begin{equation}
M_{\rm eff} = M + M_{\rm SNSI}.
\label{eq:Meff}
\end{equation}

\noindent The neutrino mass matrix ($\equiv$ $\mathcal{U^{'}} D_{\nu} \mathcal{U^{'}}^{\dagger}$) can be diagonalized by a mixing matrix $\mathcal{U^{'}} \equiv P \mathcal{U} Q^{\dagger}$. Here, Q is a Majorana rephasing matrix which can be absorbed as $Q D_\nu Q^{\dagger} = D_\nu$. The matrix P is an unphysical diagonal rephasing matrix, which can not be rotated away. We rotate P into the contribution coming from the scalar NSI and express $M_{\rm eff}$ as,

\begin{equation}
 M_{eff} \equiv \mathcal{U} D_\nu \mathcal{U}^{\dagger} + P^{\dagger} M_{SNSI} P \equiv M + \delta M 
 \label{effectiveM}.
\end{equation}

\noindent We define $\delta M$ as the perturbative term that includes the contribution of scalar NSI in which the unphysical rephasing matrix P has also been rotated into. We desire an effective and general form of $\delta M$, which will bring ease in element-wise study coupled with the neutrino mass matrix and hence we parameterize  $\delta M$ as,\\
\begin{equation}
\delta M
\equiv
\sqrt{|\Delta m^2_{31}|}
\left\lgroup
\begin{matrix}
\eta_{ee}     & \eta_{e \mu}    & \eta_{e \tau}   \\
\eta_{\mu e}  & \eta_{\mu \mu}  & \eta_{\mu \tau} \\
\eta_{\tau e} & \eta_{\tau \mu} & \eta_{\tau \tau}
\end{matrix}
\right\rgroup \,.
\label{eq:dM}
\end{equation}

\noindent We use the factor $\sqrt{|\Delta m^2_{31}|}$ as a characteristic scale.  The elements $\eta_{\alpha \beta}$ are dimensionless and they will quantify the effects of the scalar NSI.

The Hermicity of the neutrino Hamiltonian demands that the diagonal elements are real and the off-diagonal elements are complex, which may be parameterized as the following,
\begin{equation}
     \eta_{\alpha\beta}=|\eta_{\alpha\beta}|e^{i\phi_{\alpha\beta}}; \qquad  \alpha \neq \beta.
\end{equation}

\noindent For this work, we have considered a diagonal $\delta M$ which preserves the Hermicity of the Hamiltonian. This formalism would enable the exploration of the scalar NSI elements through different probability channels. The elements $\eta_{\alpha\beta}$ quantify the strength of the interactions, they can be probed in various neutrino experiments. Currently, there are not any definite bounds on these elements and we expect to impose bounds from the results of different neutrino experiments.

We have considered three cases, with one non-zero diagonal element at a time. The expressions for $M_{\rm eff}$ that we use to calculate the modified Hamiltonian in these three cases are given below,

\begin{equation}
{\rm Case~I:}~ M_{\rm eff} = \mathcal{U} {\rm diag}\left(m_1, m_2, m_3
 \right)\mathcal{U}^\dag + \sqrt{|\Delta m^2_{31}|}~ \rm diag \left( \eta_{ee}, 0, 0
 \right).
 \label{MeffCase1}
  \end{equation}
   
\begin{equation}
~~~~~~~~~{\rm Case~II:}~ M_{\rm eff} = \mathcal{U} {\rm diag}\left(m_1, m_2, m_3
 \right)\mathcal{U}^\dag + \sqrt{|\Delta m^2_{31}|}~ \rm diag \left( 0, \eta_{\mu\mu}, 0
 \right).
  \label{MeffCase2}
  \end{equation}
  
  \begin{equation}
~~~~~~~~~~~{\rm Case~III:}~ M_{\rm eff} = \mathcal{U} {\rm diag}\left(m_1, m_2, m_3
 \right)\mathcal{U}^\dag + \sqrt{|\Delta m^2_{31}|}~ \rm diag \left( 0, 0, \eta_{\tau\tau}
 \right).
  \label{MeffCase3}
  \end{equation}

    It is interesting to see that $\mathcal H_{\rm SNSI}$ has a dependence on the absolute masses of neutrinos. In this work, throughout the analysis, we have assumed the value of $m_1$ to be $10^{-5}$ eV. The values of $m_2$ and $m_3$ are accordingly obtained from the values of $\Delta m_{21}^2$ and $\Delta m_{31}^2$. In the next section, we present some probability plots and describe the simulation procedure.
   
\section{Methodology}\label{sec:Methodology}
In section~\ref{sec:oscillation probability} we have explored the effects of scalar NSI on the oscillation probabilities at DUNE. In section \ref{sec:simulation}, we have discussed the experimental details and the simulation procedure used to study the effects of scalar NSI.

\subsection{Effects on oscillation probabilities} \label{sec:oscillation probability}
The phenomena of neutrino oscillations have played a crucial role in our understanding of various properties of neutrinos from different neutrino experiments. In general, in the long baseline (LBL) neutrino experiments the most relevant neutrino oscillation channels are $\nu_\mu$ $\rightarrow$ $\nu_e$ (appearance) and $\nu_\mu$ $\rightarrow$ $\nu_\mu$ (disappearance) probability channels. We have probed the effects of scalar NSI on the neutrino oscillation probabilities.\\

\begin{table}[h]
	\centering

	\begin{tabular}{|c|c|c|}
		\hline
		Parameters & True Values\\
		\hline
		$\theta_{12}$ & 34.51$^\circ$    \\
		$\theta_{13}$ & 8.44$^\circ$   \\
		$\theta_{23}$ & 47$^\circ$   \\
		$\delta_{CP}$ & -$\pi$/2 \\
		$\Delta m_{21}^2$ & 7.56 $\times$ 10$^{-5}$ $eV^2$  \\
		$\Delta m_{31}^2$ & 2.55 $\times$ 10$^{-3}$ $eV^2$  \\
		\hline
	\end{tabular}
	\caption{The benchmark values of oscillation parameters used~\cite{NuFIT5.0}.}
	 \label{tab:mixing_parameters}
\end{table} 
We have explored the effects of the diagonal scalar NSI elements i.e. $\eta_{ee}$, $\eta_{\mu \mu}$, and $\eta_{\tau \tau}$ on oscillation probabilities. The values of the oscillation parameters used throughout the analysis are listed in table~\ref{tab:mixing_parameters}.  We see in section~\ref{sec:scalar NSI} that the Hamiltonian in the presence of the scalar NSI also depends on the absolute neutrino masses (eq.~\ref{eq:Hs}). We have assumed the value of $m_1$ to be $10^{-5}$ eV  and accordingly calculated the values of $m_2$ and $m_3$ from the values of $\Delta m_{21}^2$ and $\Delta m_{31}^2$ \cite{Aker:2021gma}. Note that, we have considered the normal ordering of neutrino mass as true hierarchy throughout the analysis. We have modified the NuOscProbExact package \cite{Bustamante:2019ggq} to include the scalar NSI effects. NuOscProbExact is a python-based neutrino oscillation probability generator that uses SU(2) and SU(3) expansions of the evolution operators to compute exact two-flavour and three-flavour neutrino oscillation probabilities for time-independent Hamiltonian. To compute the probabilities in presence of scalar NSI we have modified the Hamiltonian using the eq.~\ref{eq:Hs}. We have used the three definitions of $M_{\rm eff}$ as given is eq.~\ref{MeffCase1}, eq.~\ref{MeffCase2} and eq.~\ref{MeffCase3} as the test cases.  

\begin{figure}[h]
	\centering
	\includegraphics[width=7.5cm]{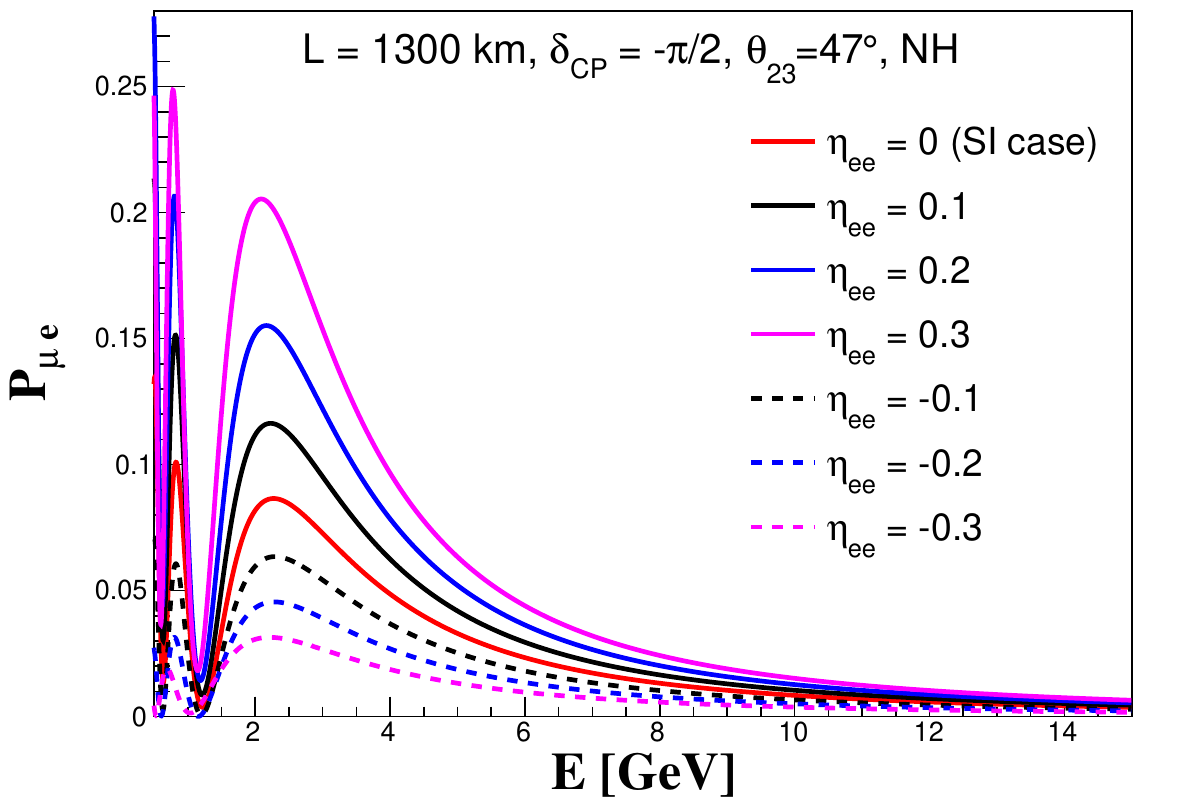}
	\includegraphics[width=7.5cm]{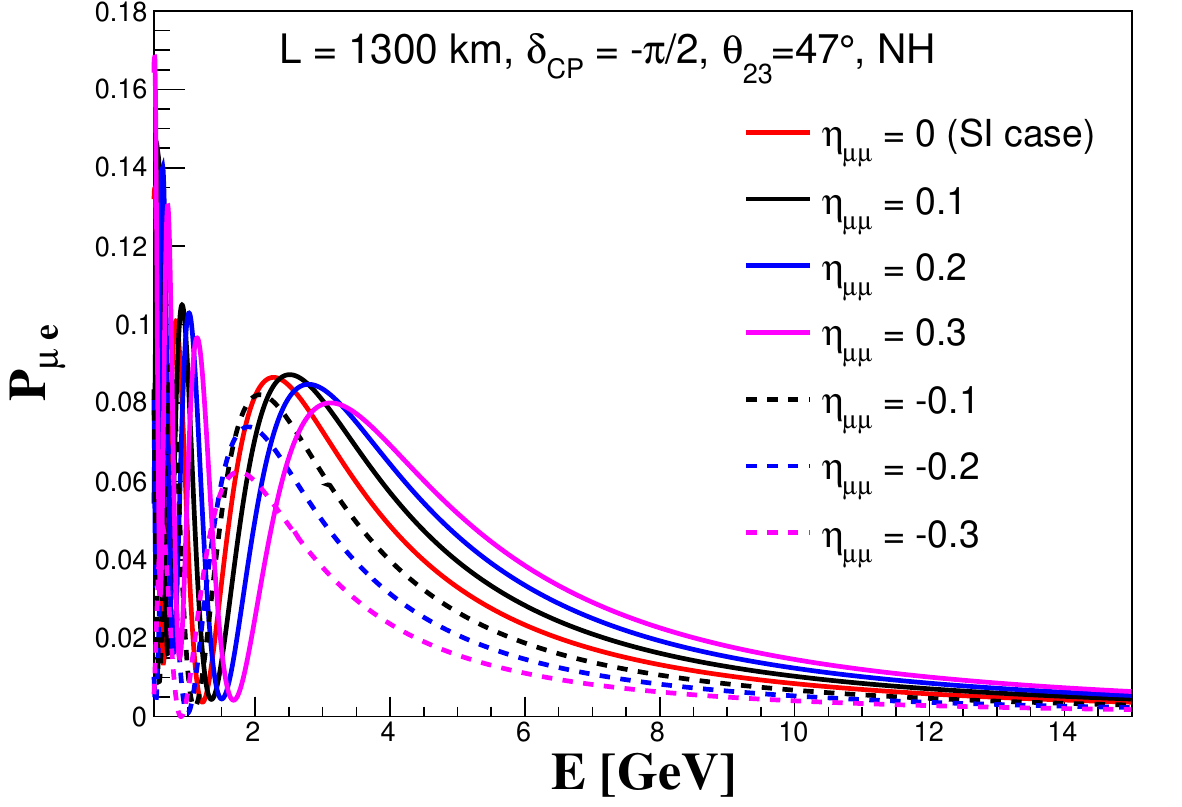}\\
	\includegraphics[width=7.5cm]{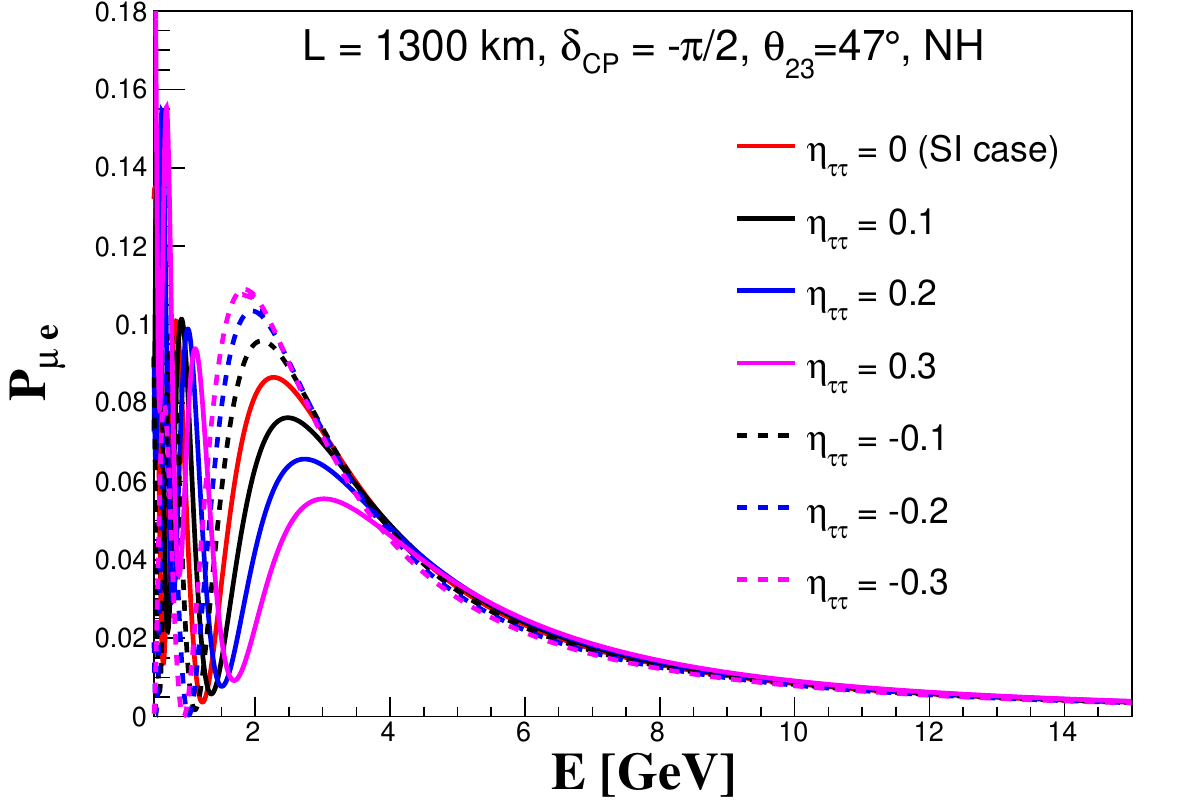}
	\caption{The effects of $\eta_{ee}$ (top-left panel), $\eta_{\mu\mu}$ (top-right panel), and $\eta_{\tau\tau}$ (bottom panel) on $P_{\mu e}$ for $\delta_{CP}$ = -$\pi$/2, $\theta_{23}$ = 47$^\circ$ and NH. In every subfigure, the red-solid curve is for the SI case and the other solid (dashed) curves are for positive (negative) non-zero values of $\eta_{\alpha\beta}$. }
	\label{fig:probability1a}
\end{figure}

In figure~\ref{fig:probability1a}, we show the appearance probability $P_{\mu e}$ for the three $M_{\rm eff}$ cases as a function of the neutrino energy at a fixed baseline of 1300 km (i.e. DUNE baseline). We have varied the energy of the neutrino beam from 0.05 GeV to 15 GeV, which is the significant energy range for DUNE. The effects of the diagonal scalar NSI elements on the appearance probability ($P_{\mu e}$) at DUNE for some chosen values of $\eta_{\alpha \beta}$ $\in$ [-0.3, 0.3] are shown. The solid-red curve, in each panel of figure~\ref{fig:probability1a}, represents the probability in the absence of scalar NSI, i.e., the standard case. The solid (dashed) black, blue, and magenta lines are for the non-zero positive (negative) choices of the scalar NSI parameter values. In the top-left panel of figure~\ref{fig:probability1a}, we show the results for case 1 (eq.~\ref{MeffCase1}), i.e., for non-zero $\eta_{ee}$. We observe that the effects of diagonal scalar NSI are reasonably significant on the oscillation probabilities of DUNE, especially around the oscillation peaks. The positive values of $\eta_{ee}$ element enhance the probabilities around the oscillation maxima while the negative values of $\eta_{ee}$ suppress the probabilities. Similarly, the top-right panel and the bottom panel in figure~\ref{fig:probability1a} show the effects of appearance probability $P_{\mu e}$ at DUNE in presence of $\eta_{\mu\mu}$ (case II, eq.~\ref{MeffCase2}) and $\eta_{\tau \tau}$ (case III, eq.~\ref{MeffCase3}) respectively. The positive (negative) values of $\eta_{\mu\mu}$ shift the oscillation peaks towards the higher (lower) energies. In addition, there is also a suppression in the probabilities as $\eta_{\mu\mu}$ shifts away from the SI case, for both positive and negative values. Also, we see that the probabilities got suppressed (enhanced) with increasing positive (negative) non-zero values of $\eta_{\tau \tau}$.

\begin{figure}[h]
	\centering
	\includegraphics[width=7.5cm]{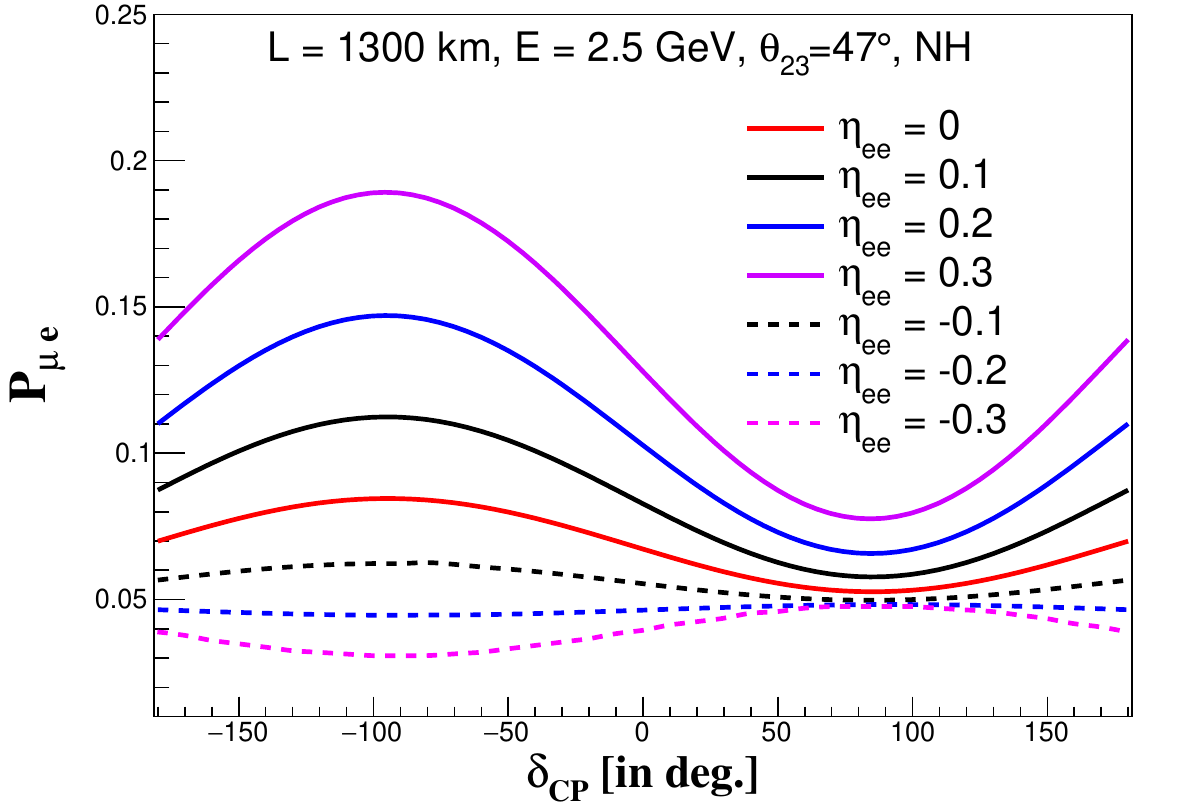}
	\includegraphics[width=7.5cm]{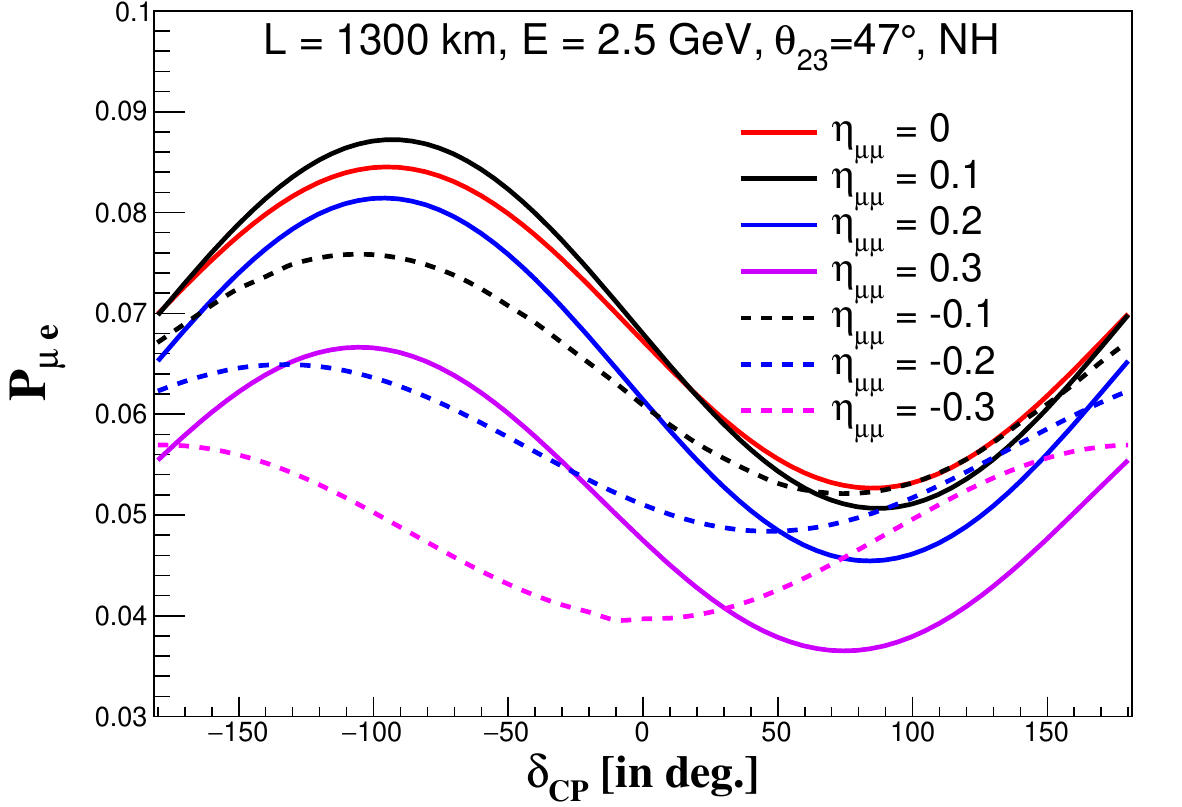}\\
	\includegraphics[width=7.5cm]{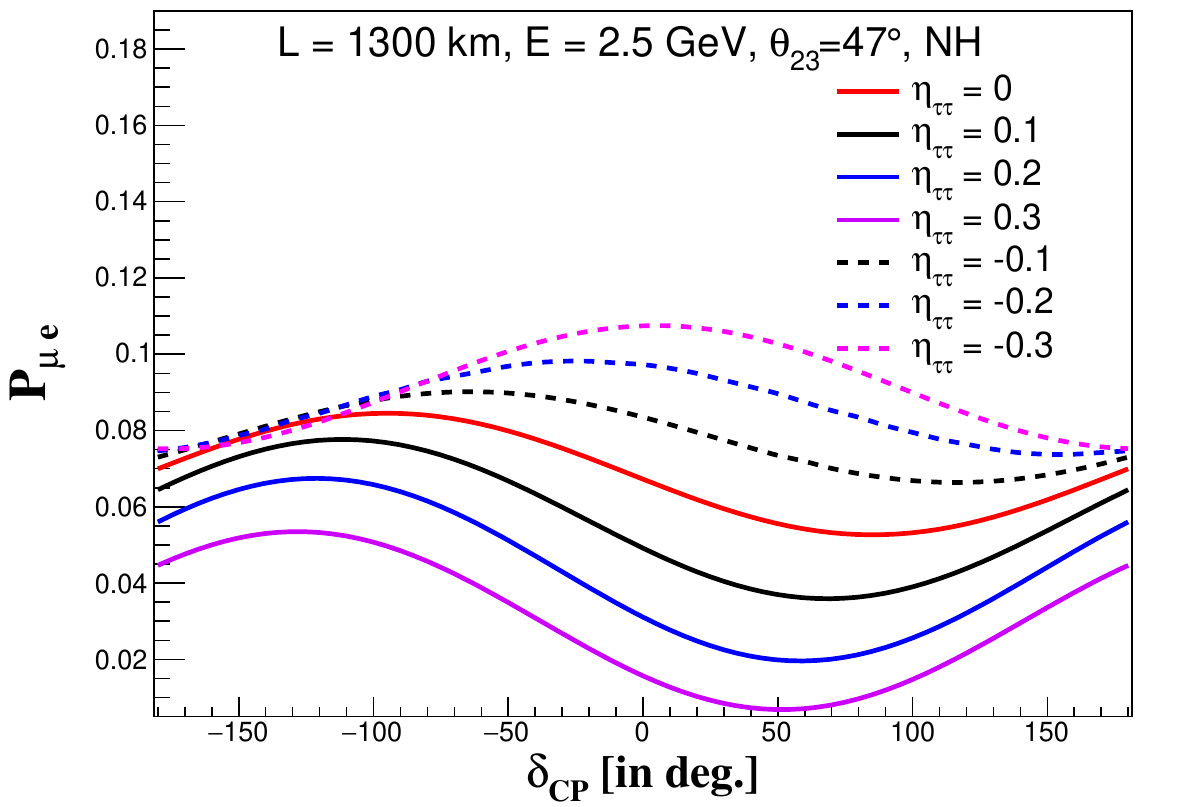}
	\caption{The effects of $\eta_{ee}$ (top-left panel), $\eta_{\mu\mu}$ (top-right panel), and $\eta_{\tau\tau}$ (bottom panel) on $P_{\mu e}$ as a function of $\delta_{CP}$ at $\theta_{23}$ = 47$^\circ$ and $E$ = 2.5 GeV. In every panel, the red-solid line is for $\eta_{\alpha\beta} = 0$ (i.e, the SI case), and other solid (dashed) coloured lines are for positive (negative) non-zero values of $\eta_{\alpha\beta}$.}
	\label{fig:probability1d}
\end{figure}

In figure~\ref{fig:probability1d}, we show the effect of scalar NSI on the appearance probability ($P_{\mu e}$) as a function of $\delta_{CP}$. We have taken the neutrino beam energy E = 2.5 GeV, $\theta_{23}$ = 47$^\circ$ and NH as the true hierarchy. In this figure, the element $\eta_{ee}$ (top-left panel), $\eta_{\mu\mu}$ (top-right panel), and $\eta_{\tau\tau}$ (bottom panel) are varied for a few chosen values in the range -0.3 to 0.3. In every panel of figure~\ref{fig:probability1d}, the red-solid line represents the SI case i.e., $\eta_{\alpha\beta}$ = 0. We see that the increasing positive (negative) values of the element $\eta_{ee}$ enhances (suppresses) the appearance probability. For non-zero $\eta_{\mu\mu}$, the probabilities are suppressed for either of the positive and the negative chosen values. For positive (negative) values of $\eta_{\tau\tau}$ the probabilities are suppressed (enhanced). It is also observed that there are some degeneracy in the probability for some values of $\eta_{\alpha\beta}$, which may impact the experiment's sensitivity towards measuring the $\delta_{CP}$ phase.

The significant dependence of the probability on $\delta_{CP}$ and $\eta_{\alpha\beta}$ is interesting and we explore it further to identify interesting regions. To compare and quantify the impact of scalar NSI on the probability, we define the parameter $\Delta P_{\alpha\beta}$ as:
\begin{equation}
\Delta P_{\alpha\beta} = P_{\alpha\beta} {\rm(with~SNSI)} - P_{\alpha\beta} {\rm(without~SNSI)}.    
\end{equation}
$\Delta P_{\alpha\beta}$ measures the change in the probability in presence of scalar NSI in comparison to the probability without scalar NSI effects.

\begin{figure}[h]
\centering
	\includegraphics[width=7.5cm]{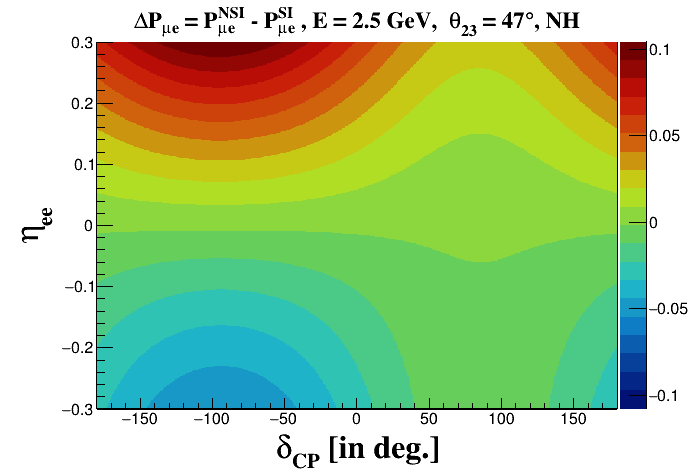}
	\includegraphics[width=7.5cm]{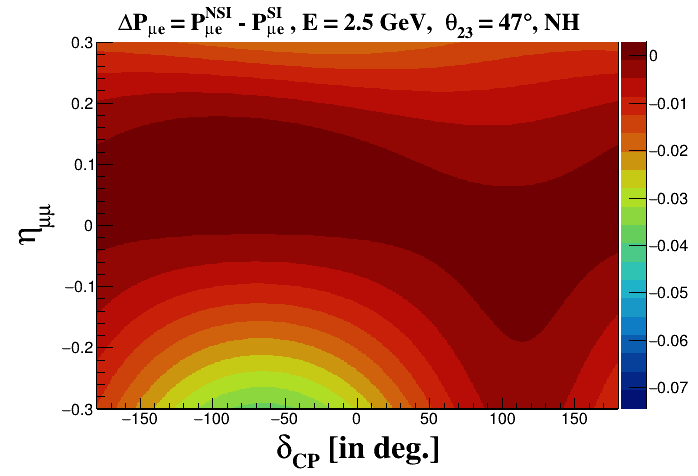}\\
    \includegraphics[width=7.5cm]{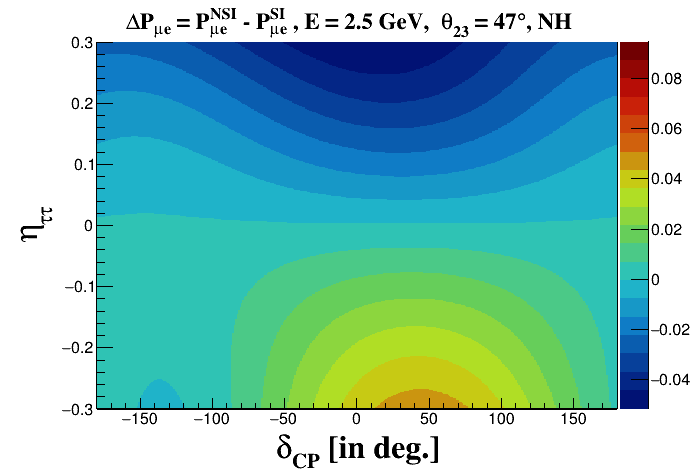}

	\caption{The variation of $\Delta P_{\alpha\beta}$ on $\eta_{\alpha\beta}$ - $\delta_{CP}$ plane. We show the results for $\eta_{ee}$ (top-left panel),  $\eta_{\mu\mu}$ (top-right panel) and   $\eta_{\tau\tau}$ (bottom panel). }
	\label{fig:dcp_2d}
\end{figure}

In figure~\ref{fig:dcp_2d}, we scan the values $\Delta P_{\alpha\beta}$ over a wide $\eta_{\alpha\beta}$--$\delta_{CP}$ range. Here we have used E = 2.5 GeV, $\theta_{23}$ = 47$^\circ$ and NH as the true hierarchy. In all the plots in figure~\ref{fig:dcp_2d}, $\eta_{\alpha\beta}$ parameters are varied in the range [-0.3, 0.3] and $\delta_{CP}$ is varied in the range [-$\pi$, $\pi$]. We see that, if the true $\delta_{CP}$ is in the range [-$\pi$, 0], a non-zero $\eta_{ee}$ can significantly impact the probability (top-left panel). The impact of $\eta_{\mu\mu}$ is slightly milder (top-right panel) while the impact of $\eta_{\tau\tau}$ is significant (bottom panel) in the $\delta_{CP}$ range [-$\pi$/3, $\pi$].

\subsection{The details of the simulation for DUNE}\label{sec:simulation}
The Deep Underground Neutrino Experiment (DUNE) \cite{DUNE1,DUNE2,DUNE3,DUNE4,DUNE5} is a proposed super-beam neutrino experiment that will be located in the USA. The beam for this experiment would be generated at Fermilab, USA and the neutrinos are expected to be detected at a distance of 1300 km from the source at Homestake Mine in South Dakota. The detector will be made up of a 35 kt - 40 kt liquid argon time projection chamber (LArTPC). The super-beam produced by Fermilab is planned to have a power of 1.2 MW - 120 GeV and will deliver 10$^{21}$ proton-on-target (POT) per year. The experiment is expected to start its data taking around 2025. These flagship configurations of this experiment make it a suitable candidate to tackle all the three unknowns of the neutrino sector i.e. the neutrino mass hierarchy, octant of $\theta_{23}$, and CP-violation in the leptonic sector.

We have used the General Long Baseline Experiment Simulator (GLoBES)~\cite{Huber:2004ka, Kopp:2006wp, Huber:2007ji}, which is a neutrino experiment simulator and is used to study the long-baseline neutrino oscillation experiments. In this work, we have considered a liquid-argon detector with a baseline of 1300 km, which corresponds to the specifications of DUNE. We have considered a run time for 5 years of neutrino and 5 years for antineutrino mode to simulate the experiment, which gives a total exposure of 35 $\times$ 10$^{22}$ kt-POT-yr. We have used the combined appearance and disappearance channel in our analysis. The signal normalization for neutrino (antineutrino) mode is taken as 2\% (5\%). The background normalization error is taken as 10\% for both of the modes. The energy resolution for $\mu$ ($R_\mu$) and e ($R_e$) are taken to be 20\%/$\sqrt{E}$ and 15\%/$\sqrt{E}$ respectively. The energy calibration error for both the neutrino and antineutrino mode is taken as 5\%. We have used the Technical Design Report (TDR) of DUNE~\cite{DUNE3} for background and choice of systematics. The details of the experimental configuration and the systematic uncertainties taken in our simulations are listed in table~\ref{Table 2a}.\\

\begin{table}[!h]
\centering
\begin{tabular}{ |l| l l |ll| }
\hline
Detector details & \multicolumn{2}{c |}{Normalisation error} & \multicolumn{2}{c |}{Energy calibration error} \\
                                         \cline{2-5}
                                        & Signal & Background       & Signal & Background                              
                                        \\ 
\hline
Baseline = 1300 km 
&&&&\\
 Runtime (yr) = 5 $\nu$ + 5 $\bar \nu$ 
  & $\nu_e : 5\%$  & $\nu_e : 10\%$ &   $\nu_e : 5\%$ & $\nu_e : 5\%$\\
35 kton, LArTPC & & &&\\
$\varepsilon_{app}=80\%$, $\varepsilon_{dis}=85\%$
& $\nu_\mu : 5\%$ & $\nu_\mu : 10\%$ & $\nu_\mu : 5\%$ & $\nu_\mu : 5\%$\\
{{$R_e=0.15/{\sqrt E}$, $R_\mu=0.20/{\sqrt E}$}} &&&&\\
\hline
\end{tabular}
\caption{\label{Table 2a} 
Details of detector configurations, efficiencies, resolutions, and systematic uncertainties for DUNE. Here, $\varepsilon_{app}$ and $\varepsilon_{dis}$ are signal efficiencies for $\nu_{e}^{CC}$ and $\nu_{\mu}^{CC}$ respectively. Also, $R_e$ and $R_\mu$ is energy resolutions for signal $\nu_{e}^{CC}$ and $\nu_{\mu}^{CC}$ respectively.}
\end{table}

\section{Results and Discussion}\label{sec:results}

\begin{figure}[h]
	\centering
	\includegraphics[width=9cm]{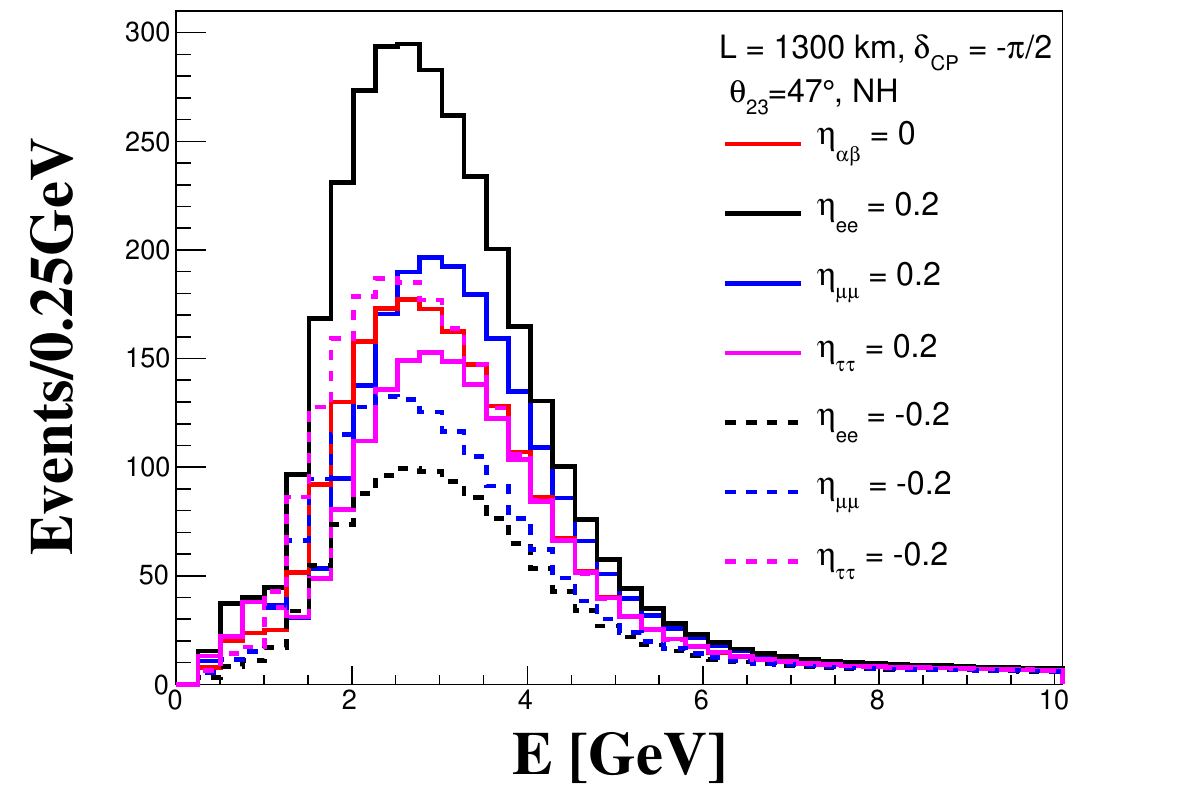}
	\caption{The binned event rates as a function of the neutrino energy for different choices of $\eta_{\alpha\beta}$. The results shown here are for $\delta_{CP}$ = -$\pi$/2, $\theta_{23}$ = 47$^\circ$ and NH. }
	\label{fig:event_rate}
\end{figure}

In order to explore the sensitivity of DUNE in the presence of scalar NSI, we first look into the event rates at the detector. In figure~\ref{fig:event_rate}, we show the binned events as a function of the reconstructed neutrino energy. The values of the mixing parameters used to generate the event plots are listed in table~\ref{tab:mixing_parameters}. The solid-red histogram represents the event rate in absence of scalar NSI. The solid (dashed) histograms represent the binned events in presence of positive (negative) scalar NSI elements viz. $\eta_{ee}$, $\eta_{\mu\mu}$, and $\eta_{\tau\tau}$. Note that the behavior of the binned events are in good agreement with the probabilities. In presence of non-zero positive (negative) $\eta_{ee}$, the number of events increases (decreases) compared to the `no scalar NSI' case for each bin. Also for positive (negative) values of $\eta_{\mu\mu}$ and $\eta_{\tau\tau}$ the peaks of the event plots shifts towards the higher (lower) energy bin. In addition, for positive (negative) values of $\eta_{\mu\mu}$ the number of events for each bin increases (decreases) around the oscillation maxima, whereas for positive (negative) values of $\eta_{\tau\tau}$ the number of events for each bin decreases (increases) around the oscillation maxima. 

We define the statistical $\chi^2$ to probe whether an experiment can distinguish between CP-conserving ( $\delta_{CP}$ = 0, $\pm$ $\pi$) and CP-violating values ( $\delta_{CP}$ $\neq$ 0, $\pm$ $\pi$) as follows,

\begin{equation}
\label{eq:chisq}
\chi^2 \equiv  \min_{\eta}  \sum_{i} \sum_{j}
\frac{\left[N_{true}^{i,j} - N_{test}^{i,j} \right]^2 }{N_{true}^{i,j}},
\end{equation}\\

\noindent where, $N_{true}^{i,j}$ and $N_{test}^{i,j}$ are the numbers of true and test events in the $\{i,j\}$-th bin respectively. The significance is denoted by $n\sigma$, where $n \equiv \sqrt{\Delta \chi^2}$. 

Our prime goal in this analysis is to point out the effects of diagonal scalar NSI elements on the CP-violation measurements at DUNE. We have studied CP-violation as well as CP-precision measurement capabilities of the experiment in presence of scalar NSI. The capability of an experiment to differentiate between CP-conserving and CP-violating values of $\delta_{CP}$ is a measure of its CP sensitivity. We have considered normal hierarchy (NH) as the true hierarchy and higher octant (HO) as the true octant (true $\theta_{23}$ = 47$^\circ$) throughout the analysis unless otherwise specified. We have marginalized over the systematic uncertainties. Note that, the mixing parameter values used in the analysis are taken from table \ref{tab:mixing_parameters} unless otherwise mentioned.

\begin{figure}[h]
	\centering
	\includegraphics[width=7cm]{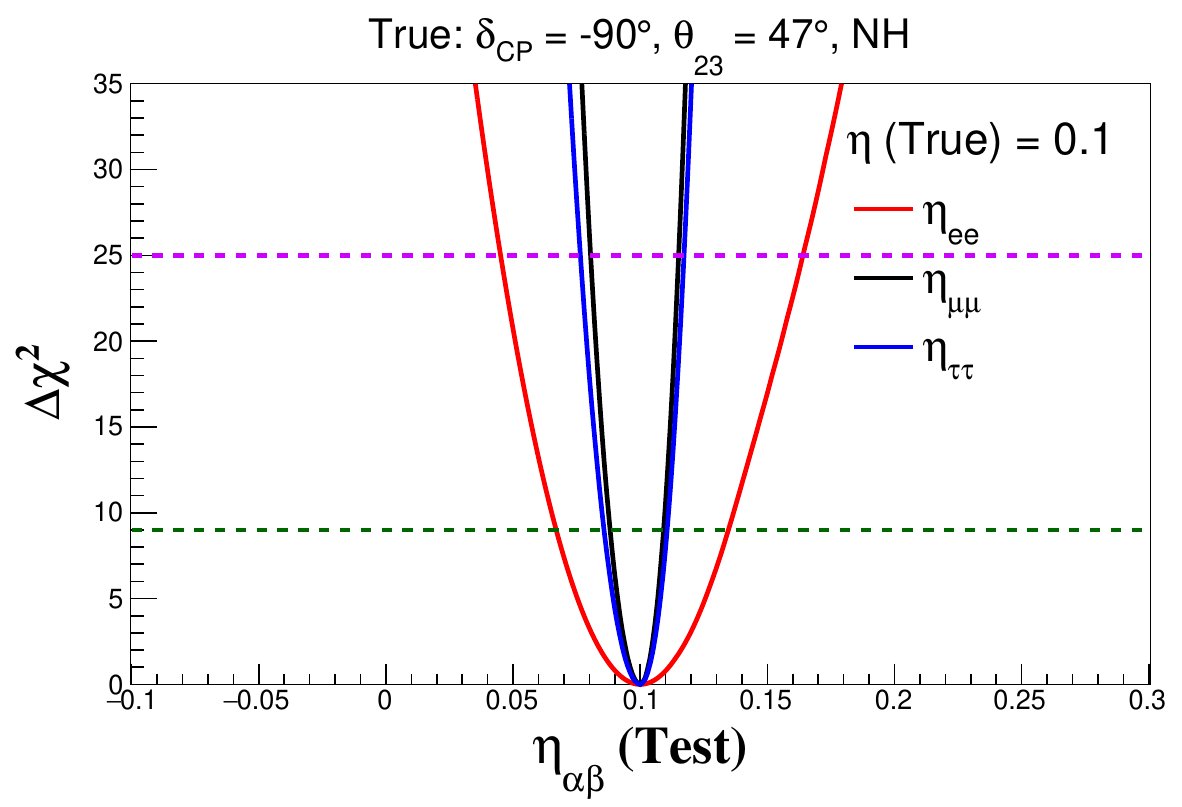}
	\includegraphics[width=7cm]{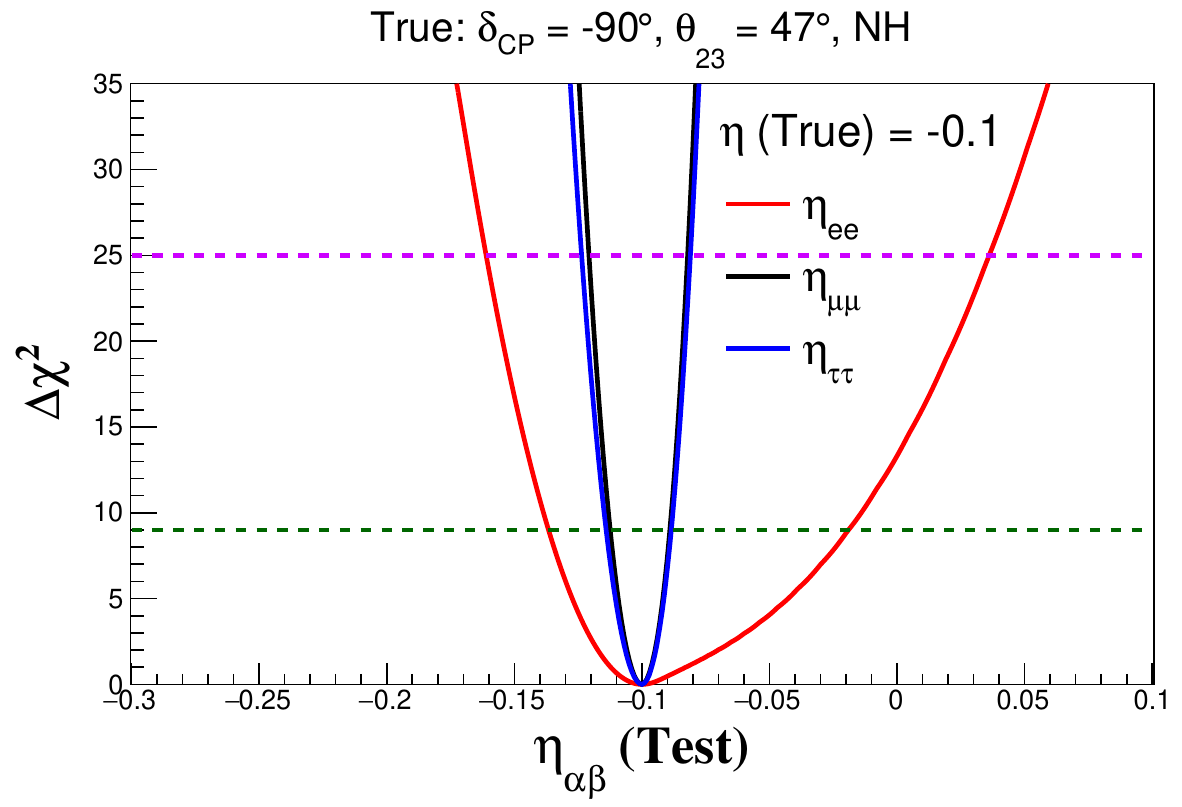}\\
	\caption{The sensitivity of DUNE to true $\eta_{\alpha\beta}$ = 0.1 (left panel) and true $\eta_{\alpha\beta}$ = -0.1 (right panel) at true $\delta_{CP}$ = -$\pi$/2 and true $\theta_{23}$ = 47$^\circ$. In both of the plots, the red-solid curve is for $\eta_{ee}$, the black-solid curve is for $\eta_{\mu\mu}$, and the blue-solid curve is for $\eta_{\tau\tau}$.}
	\label{fig:chi2_1}
\end{figure}

 In figure~\ref{fig:chi2_1}, we show the sensitivity of DUNE towards $\eta_{\alpha\beta}$ where we take the true $\eta_{\alpha\beta}$ as 0.1 (left panel) and -0.1 (right panel) and vary the test values of $\eta_{\alpha\beta}$ in the range -0.3 to 0.3. We see that, in both of the cases, the experiment's capability towards constraining $\eta_{ee}$ is milder as compared to that towards $\eta_{\mu\mu}$ and $\eta_{\tau\tau}$.

\begin{figure}[h]
	\centering
	\includegraphics[width=7.5cm]{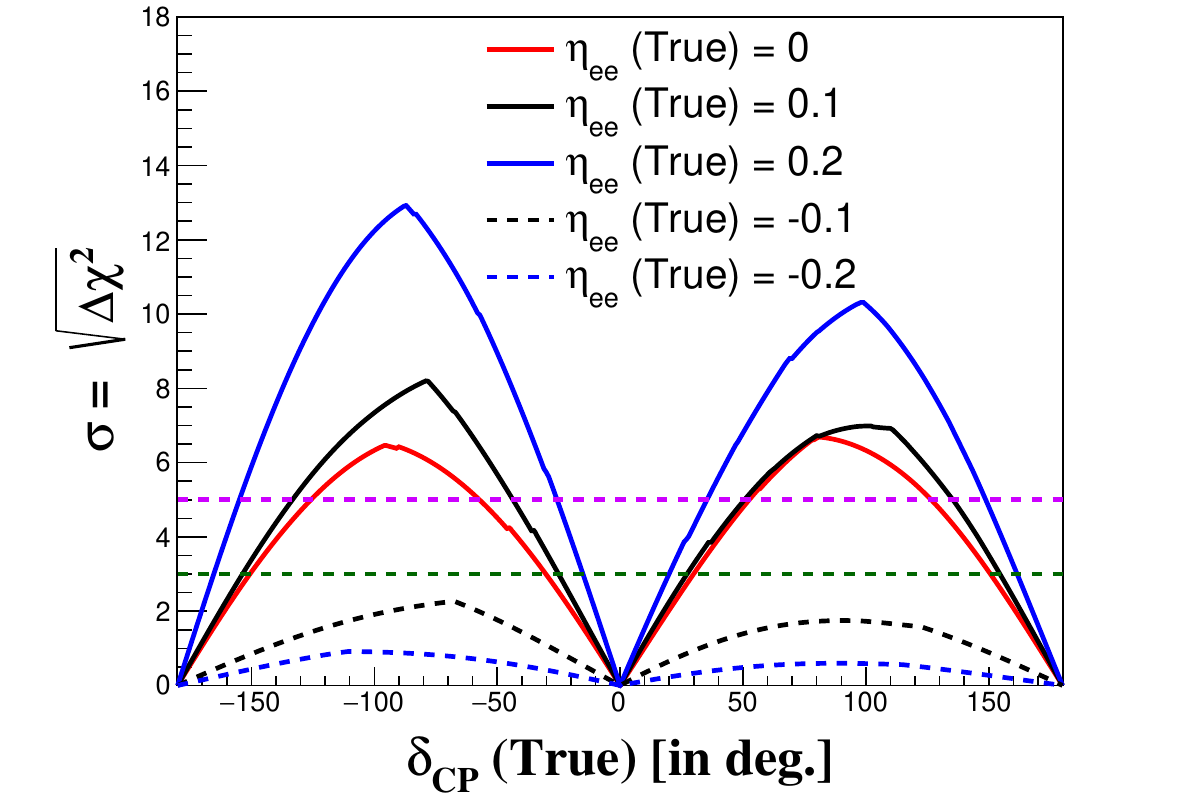}
	\includegraphics[width=7.5cm]{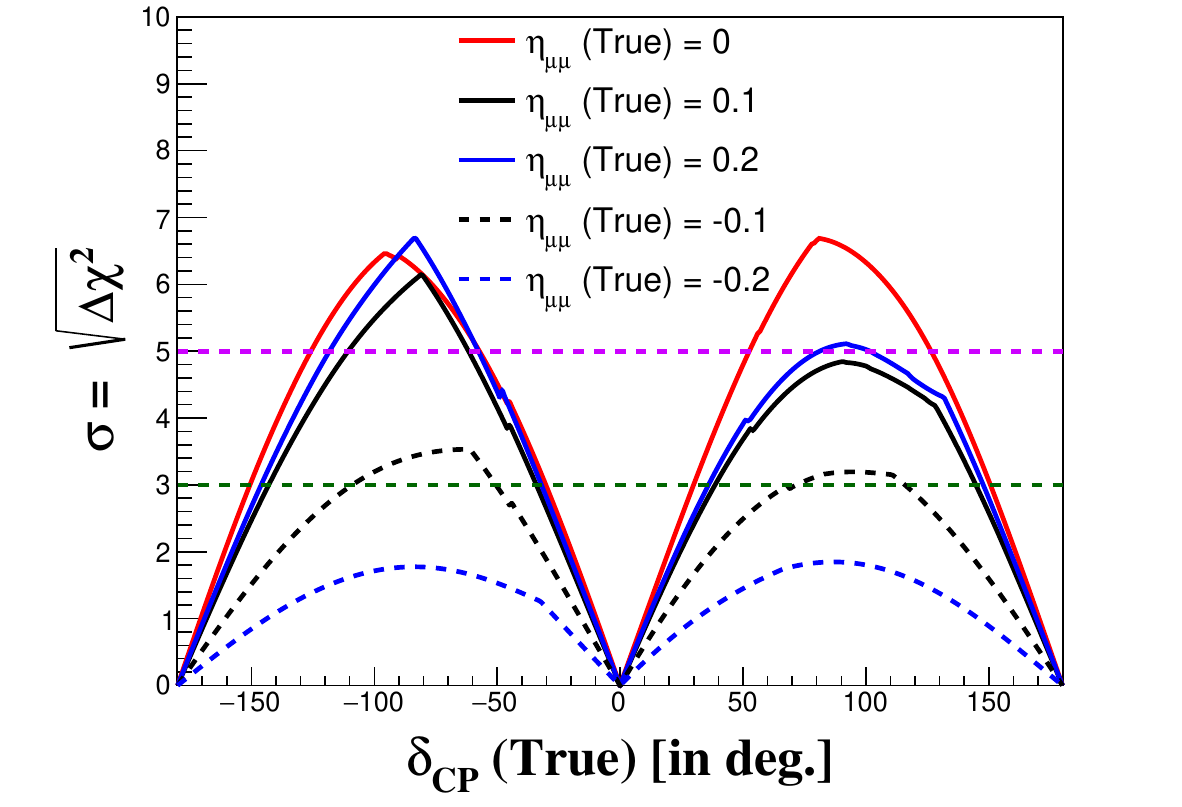}\\
	\includegraphics[width=7.5cm]{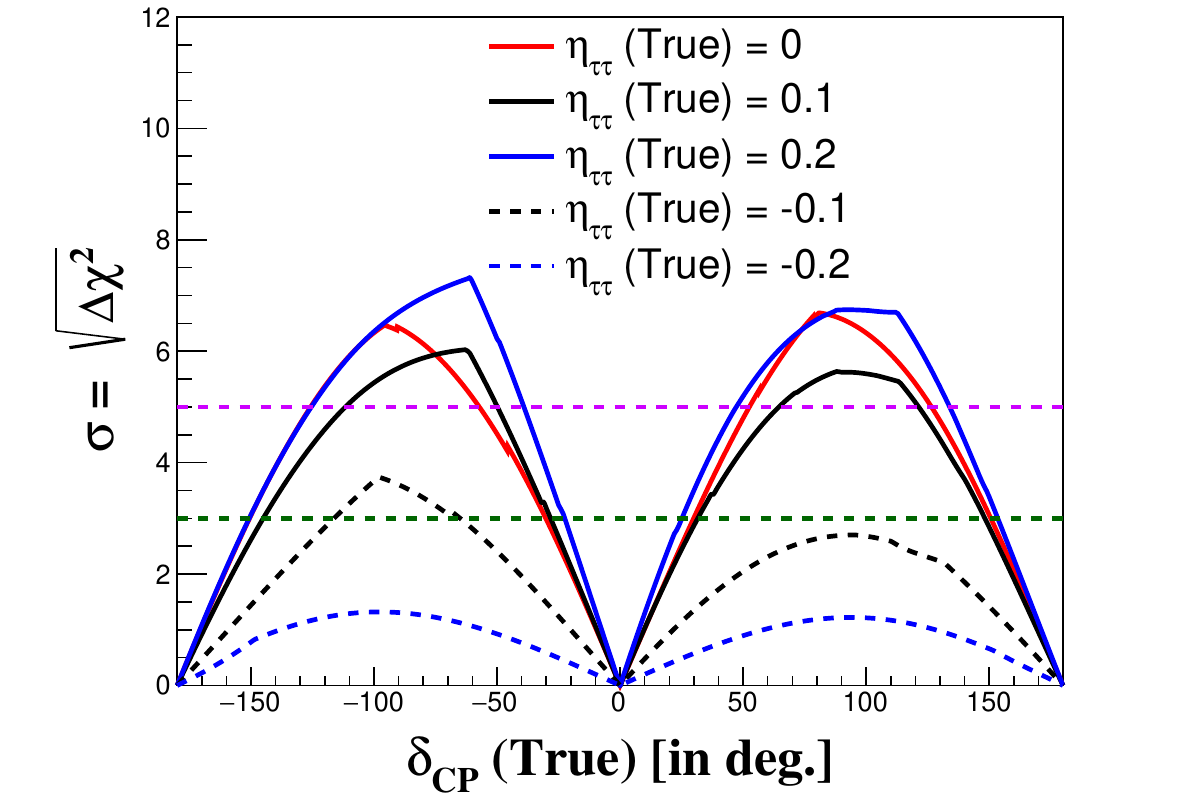}
	\caption{The CP-violation sensitivity of DUNE in presence
		of scalar NSI element $\eta_{\alpha\beta}$. The results for $\eta_{ee}$ (top-left panel), $\eta_{\mu\mu}$ (top-right panel), and $\eta_{\tau\tau}$ (bottom panel) are shown. In all the three plots, the red curve is for the `no scalar NSI' case whereas the black and the blue solid (dashed) curves are for the chosen positive (negative) non-zero $\eta_{\alpha\beta}$.}
	\label{fig:chi2_2}
\end{figure}

In figure~\ref{fig:chi2_2}, we show the CP-violation (CPV) sensitivity of DUNE in the presence of the scalar NSI elements, $\eta_{ee}$, $\eta_{\mu\mu}$, and $\eta_{\tau\tau}$. To perform this analysis, we have excluded CP non-conserving values in the test $\delta_{CP}$ range. We have varied the true value for $\delta_{CP}$ in the allowed range [-$\pi$,$\pi$]. The CPV sensitivity is obtained by calculating the minimum of $\Delta \chi^2$, after marginalizing over the allowed 3$\sigma$ range of $\theta_{23}$ and $\Delta m^2_{31}$. We define the CPV sensitivity as,

\begin{equation}
{\Delta \chi}^{2}_{\rm CPV}~(\delta^{\rm true}_{\rm CP}) = {\rm min}~\left[\chi^2~(\delta^\text{true}_{CP},\delta^\text{test}_{CP}=0),~\chi^2 (\delta^\text{true}_{CP},\delta^\text{test}_{CP}=\pm \pi)\right ].
\end{equation}

\noindent Here we have also marginalized over the test values of $\eta_{\alpha\beta}$ for the NSI cases. In all the three panels of figure~\ref{fig:chi2_2}, the red-solid curve represents the case with `no scalar NSI'. The black and blue lines (solid-dashed) represent the cases with non-zero positive (negative) values of the chosen scalar NSI parameters. The horizontal magenta and green dashed lines show the 5$\sigma$ and 3$\sigma$ CL respectively. It may be observed from figure~\ref{fig:chi2_2} (top-left panel) that, the CPV sensitivity of the experiment is significantly enhanced (suppressed) in presence of the positive (negative) $\eta_{ee}$. For the chosen values of negative $\eta_{ee}$, the sensitivity lies below the reach of 3$\sigma$ CL, and it will deteriorate the CPV sensitivity of DUNE. In figure~\ref{fig:chi2_2} (top-right panel), it is observed that for the chosen positive values of $\eta_{\mu\mu}$ the sensitivity of DUNE is not much altered in the negative half-plane of $\delta_{CP}$. In comparison, the positive $\eta_{\mu\mu}$ deteriorates the CPV sensitivity from the `no scalar NSI' case in the positive half-plane of $\delta_{CP}$.  For the chosen negative non-zero values of $\eta_{\mu\mu}$, the sensitivity gets significantly suppressed compared to the `no scalar NSI' sensitivity. Note that for the positive $\eta_{\mu\mu}$ value the sensitivities, without and with scalar NSI, almost overlap in the region $\sim$ [-90$^\circ$, 0$^\circ$]. That is why in this region it is difficult to pinpoint whether the CP sensitivity is faked by the scalar NSI effects. We show that if scalar NSI exists, then CP sensitivity measured at the DUNE far detector will get affected.

\begin{figure}[h]
	\centering
	\includegraphics[width=7.5cm]{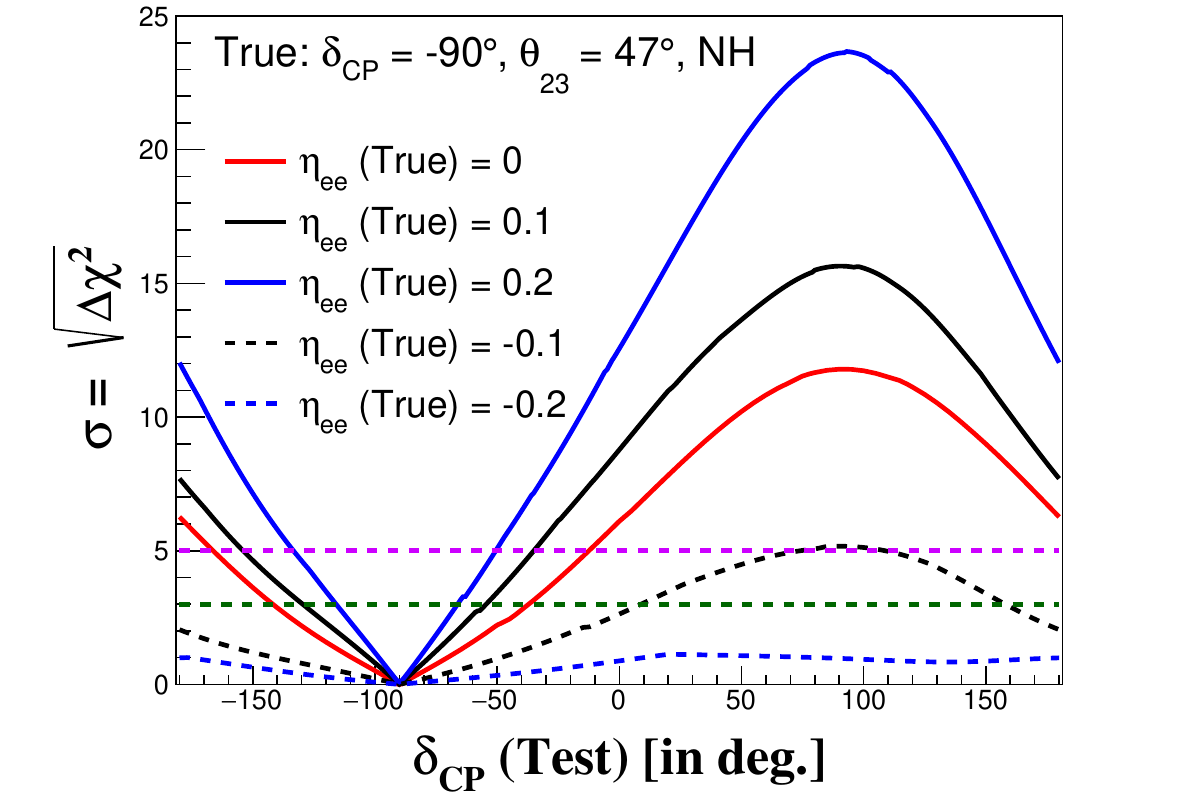}
	\includegraphics[width=7.5cm]{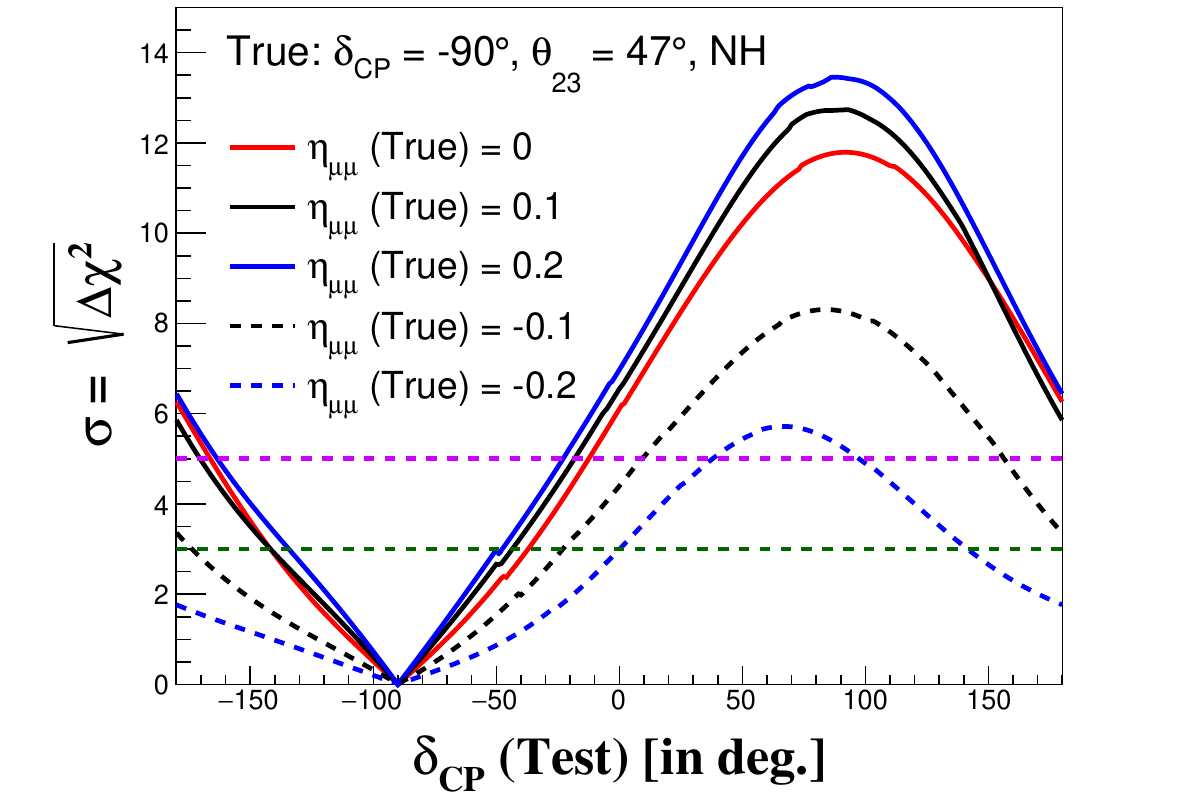}\\
	\includegraphics[width=7.5cm]{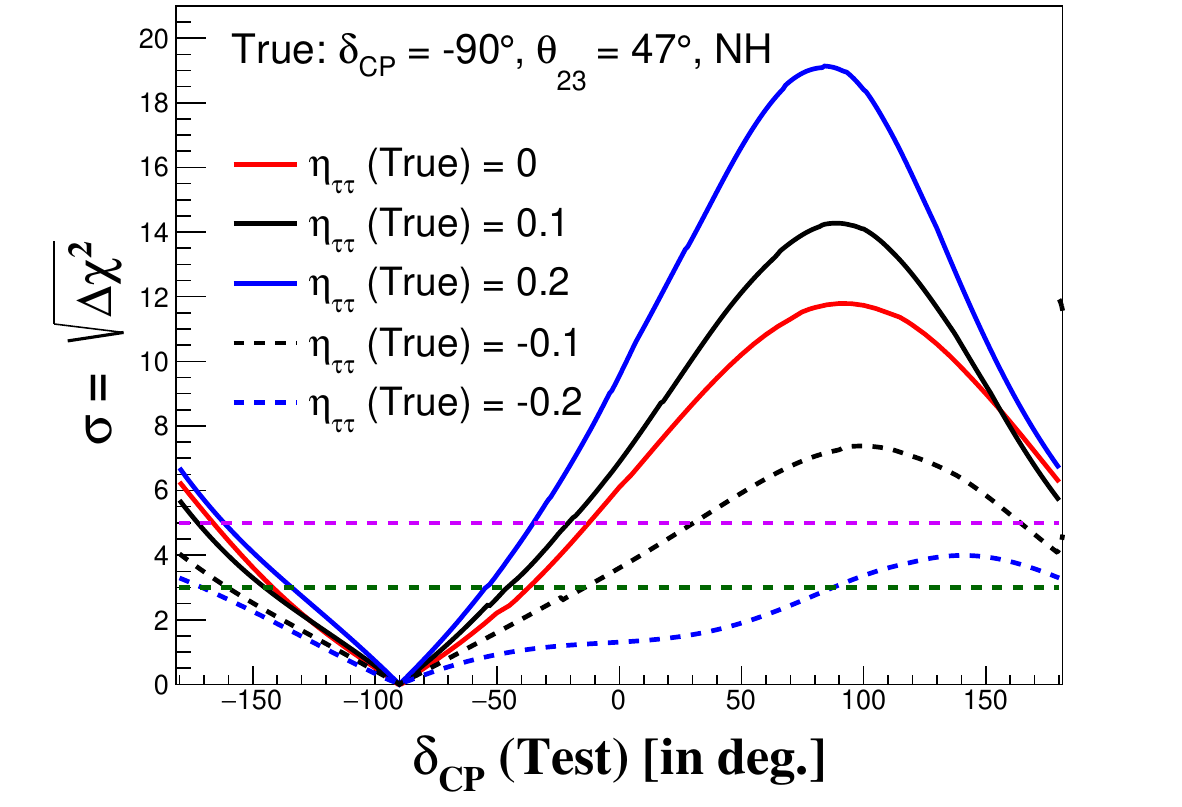}\\
	\caption{The CP-precision sensitivity in presence of $\eta_{\alpha\beta}$ for true $\delta_{CP}$ = -$\pi$/2 and true $\theta_{23}$ = 47$^\circ$. We show the results for $\eta_{ee}$ (top-left panel), $\eta_{\mu\mu}$ (top-right panel) and $\eta_{\tau\tau}$ (bottom panel). In all the three plots, the red line is for the SI case.}
	\label{fig:cp_precision_ee}
\end{figure}

In figure~\ref{fig:cp_precision_ee}, we show the CP-precision measurement capability of DUNE in presence of diagonal scalar NSI elements $\eta_{ee}$, $\eta_{\mu\mu}$, and $\eta_{\tau\tau}$. We have fixed all the true values of the mixing parameters and marginalized the test $\delta_{CP}$ and $\theta_{23}$ over the allowed ranges i.e [-$\pi$, $\pi$] and [$40^\circ$, $50^\circ$] respectively. We have considered the true $\delta_{CP}$ to be -$90^\circ$.  We have also marginalized the test $\eta_{\alpha\beta}$ from -0.2 to 0.2 and we have plotted $\sigma$ (= $\sqrt{\Delta\chi^2}$) as a function of the test values of $\delta_{CP}$. The magenta-dotted and green-dotted lines represent the $\Delta\chi^2$ corresponding to 5$\sigma$ and 3$\sigma$ CL respectively. The analysis signifies, knowing the true values of $\delta_{CP}$, how precisely DUNE can constrain the test $\delta_{CP}$ values. 

From figure~\ref{fig:cp_precision_ee} (top-left panel), it may be observed that the capability of DUNE in constraining the $\delta_{CP}$ phase gets enhanced (worsened) for positive (negative) non-zero values of $\eta_{ee}$. For the `no scalar NSI' case (the red curve), DUNE should be able to measure the phase $\delta_{CP}$ with a precision of $\sim$ -$90^{\circ^+45^\circ}_{-48^\circ}$ at 3$\sigma$ CL. However, in the presence of scalar NSI, the precision may get either improved or worsened. For example, for $\eta_{ee}$ = 0.2 the experiment's capability in constraining $\delta_{CP}$ improves by -$90^{\circ^+40^\circ}_{-30^\circ}$. For $\eta_{ee} \leq$ -0.1, its CP-precision measurement ability lies below 5$\sigma$. 

As shown in the top-right panel of figure~\ref{fig:cp_precision_ee}, the presence of $\eta_{\mu\mu}$ too affects the sensitivities in constraining the $\delta_{CP}$ values. For the `no scalar NSI' case (the red curve) the ability of DUNE in constraining the $\delta_{CP}$ phase is of the order $\sim$ -$90^{\circ^+45^\circ}_{-48^\circ}$. In addition, for all the non-zero negative values of $\eta_{\mu\mu}$ the ability to constrain the $\delta_{CP}$ phase gets worsened, whereas the positive $\eta_{\mu\mu}$ makes a marginal improvement in constraining the $\delta_{CP}$ phase. For example, with $\eta_{\mu\mu}$ = -0.2 (the blue dashed line), the precision of $\delta_{CP}$ goes below the 3$\sigma$ CL as compared to the SI case. We also observe in figure~\ref{fig:cp_precision_ee} (bottom panel) that the effect of $\eta_{\tau\tau}$ on the CP-precision sensitivity is similar to that of $\eta_{ee}$. Note that, a positive $\eta_{\tau\tau}$ improves the precision measurement capability while a negative $\eta_{\tau\tau}$ reduces the CP-precision sensitivity.
 
\section{Summary and Concluding Remarks}\label{sec:conclusion}

In this precision era of neutrino physics, it is very crucial to identify the subdominant effects like NSI in the neutrino experiments and their effects on the physics potential of different experiments. In this work, we have primarily taken DUNE as an LBL candidate to study the effects of scalar NSI. We have explored the impact of the diagonal elements of the scalar NSI matrix on the experiment by performing a $\chi^2$ analysis. The experiment's sensitivity to $\eta_{\mu\mu}$ and $\eta_{\tau \tau}$ are close to each other with marginally better sensitivity for $\eta_{\mu\mu}$. We have observed that the sensitivity toward constraining $\eta_{ee}$ is slightly lower.   We have also performed a CP-violation sensitivity study for DUNE in presence of these scalar NSI elements. We have seen that the inclusion of scalar NSI can significantly impact the CP sensitivities at DUNE. For example, for a possible negative value of $\eta_{ee}$ like -0.10, the CP sensitivity of DUNE lies below the 3$\sigma$ CL. However, for the positive non-zero values of $\eta_{ee}$, the CP sensitivity of DUNE mostly gets enhanced as compared to the `no scalar NSI' case. Hence, the effects of scalar NSI can not be ignored in the LBL experiments like DUNE, given the remarkable accuracy and precision provided by the experiment. Similarly from our CP-precision study, we have observed that the ability of DUNE in constraining the CP phase gets a significant impact in presence of scalar NSI. The presence of positive $\eta_{\alpha\beta}$ elements mostly improves the ability to measure the $\delta_{CP}$, whereas negative $\eta_{\alpha\beta}$ elements mostly reduce the ability of the experiment in constraining $\delta_{CP}$ values. It is extremely crucial to put some constraints on these scalar NSI parameters for the correct interpretation of data from various neutrino experiments. In addition, it may also be very interesting to probe scalar NSI to various neutrino mass models as it directly affects the mass term in the neutrino Hamiltonian.

\section*{Acknowledgment}
The work is supported by the Research and Innovation grant 2021 (DoRD/RIG/10-73/ 1592-A), Tezpur University, received by AM. MMD acknowledges the Science and Engineering Research Board (SERB), DST for the grant EMR/2021/002961. The authors also acknowledge the DST FIST grant SR/FST/PSI-211/2016(C) received by the Department of Physics, Tezpur University.


\bibliographystyle{JHEP}
\bibliography{scalar_NSI}
\end{document}